# Tripartite entanglement transfer from flying modes to localized qubits

Federico Casagrande,[1,2,*] Alfredo Lulli,[1,2,†] and Matteo G.A. Paris[1,2,3,‡]

[1]*Dipartimento di Fisica dell'Università di Milano, Italia*
[2]*CNISM, UdR Milano, I-20133 Milano, Italia*
[3]*I.S.I. Foundation, I-10133 Torino, Italia*
(Dated: November 4, 2008)

We investigate the process of entanglement transfer from a three-mode quantized field to a system of three spatially separated qubits each one made of a two-level atom resonantly coupled to a cavity mode. The optimal conditions for entanglement transfer, evaluated by atomic tripartite negativity, are derived for radiation prepared in qubit-like and Gaussian entangled states in terms of field parameters, atom-cavity interaction time, cavity mirror losses, and atomic preparation. For qubit-like states we found that for negligible cavity losses some states may completely transfer their entanglement to the atoms and/or be exactly mapped to the atomic state, whereas for Gaussian states we found a range of field parameters to obtain a large entanglement transfer. The purity of the three-qubit states and the entanglement of two-qubit subsystems are also discussed in some details.

## I. INTRODUCTION

Physical systems of interest for quantum information processing (QIP) have been mostly those where entanglement is present at the stationary state or may be created by means of some achievable interaction Hamiltonian. In turn, optical systems have been a privileged framework for encoding and manipulating quantum information, since bipartite and multipartite entanglement may be effectively generated either in the discrete or continuous variable (CV) regime. On the other hand, the development of QIP also requires localized registers, e.g. for the storage of entanglement in quantum memories. Moreover, effective protocols for the distribution of entanglement would allow one to realize quantum cryptography over long distances [1], as well as distributed quantum computation [2] and distributed network for quantum communication purposes [3]. Multiphoton states might be optimal when considering long distance communication, where they may travel through free space or optical fibers exploiting the robustness of their entanglement against losses, for example in quantum teleportation with noisy channels [4–9]. Recently, the bipartite process of entanglement transfer from a free propagating quantized light to an atomic system has been widely investigated[10–19] and achieved experimentally [20–23].

The natural extension of these studies concerns multipartite entanglement, whose structure is currently under investigation mainly in the case of mixed states. Quantum correlations in the multipartite systems have a much richer structure than in the bipartite case [24, 25], and may be used to implement improved information processing and distributed quantum computing, as well as to reveal higher nonlocality features of quantum mechanics. In particular, tripartite entanglement is a resource to increase security of quantum cryptography [26] and it finds applications in quantum secret sharing [27] and quantum cloning [28]. It also provides a mean to implement transfer of quantum information without any classical channel [29]. In turn, the generation of tripartite entanglement for qubits has been analyzed for several physical systems including cavity quantum electrodynamics [30, 31], as well as trapped ion quantum computer [32] and magnetic systems in a ring geometry [33]. In the framework of CV systems tripartite entanglement have been theoretically investigated and experimentally realized firstly by combining on beam splitters the two-mode squeezed states emitted by optical parametric amplifiers (OPA)[34–37]. Thus the generation of multipartite entanglement only by a nonlinear optical process has been proposed [38, 39] allowing entangled modes with different frequencies. Very recently, generation of tripartite entangled CV fields by means of type-II second-order harmonic generation with a triply resonant optical cavity below threshold has been discussed [40].

In this paper we extensively investigate the process of entanglement transfer between flying radiation and qubits in the case of tripartite systems. In particular, we analyze the resonant interaction in the strong coupling regime of a three-mode quantized field with a system of three localized and spatially separated qubits each one isolated by a local environment, generalizing our recent work on bipartite systems [18]. As we will see, transfer of entanglement may be effectively achieved for some interaction times, especially for CV fields that can be well approximated by qubit-like states also in the case of mixed state preparations. We also investigate the effect of losses in the coupling between the CV field modes and the qubit local environments as well as the effect of different atomic preparations.

[*]Electronic address: federico.casagrande@mi.infn.it
[†]Electronic address: alfredo.lulli@unimi.it
[‡]Electronic address: matteo.paris@fisica.unimi.it

The paper is structured as follows. In Section II we describe in detail the entanglement transfer model and briefly review the tripartite negativity that we choose to estimate the final three-qubits entanglement. In Section III we deal with the case of field prepared in qubit-like entangled states with only few Fock components excited, whereas the case of an experimentally feasible CV Gaussian entangled state is analyzed in Section IV. The effects of different preparation of the qubits is analyzed in Section V. Section VI closes the paper with some concluding remarks.

## II. THE ENTANGLEMENT TRANSFER MODEL

We consider the entanglement transfer process from a three-mode CV field and a system of three localized and spatially separated qubits, each one interacting resonantly with one mode by means of a local environment. The scheme is analogous to that we have recently investigated to describe the entanglement transfer in the case of bipartite systems [18]. Here we generalize the model to include the field prepared also in a mixed state. We assume that each mode of the CV field is first injected into a cavity and then interacts resonantly with a two-level atom by Jaynes-Cummings (JC) interaction [41] (see Fig. 1). For the CV field we consider a general mixed state written in the Fock number basis $|pqr\rangle_f = \{|p\rangle_1 \otimes |q\rangle_2 \otimes |r\rangle_3\}_{p,q,r=0}^{\infty}$ as:

$$\hat{\rho}_f(\vec{x}) = \sum_{p,q,r,p',q',r'=0}^{\infty} a_{p,q,r,p',q',r'}(\vec{x})|pqr\rangle_f\langle p'q'r'|, \quad (1)$$

where the complex coefficients $a_{p,q,r,p',q',r'}(\vec{x}) = \langle pqr|\hat{\rho}_f(\vec{x})|p'q'r'\rangle$ satisfy a normalization condition and $\vec{x}$ is a vector of parameters to characterize the three-mode field state.

A convenient way to describe the injection of each mode into the corresponding cavity is to use a simple linear coupling, neglecting the cavity mode dissipation. Under this assumptions (beam-splitter approach), the resonant cavity-CV field mode coupling is described by the unitary operator:

$$\hat{B}_\alpha(\theta) = \exp[-\theta(\hat{f}_\alpha^\dagger \hat{c}_\alpha - \hat{f}_\alpha \hat{c}_\alpha^\dagger)] \quad (2)$$

where $\hat{f}_\alpha$ ($\hat{f}_\alpha^\dagger$) and $\hat{c}_\alpha$ ($\hat{c}_\alpha^\dagger$) ($\alpha = \{A, B, C\}$) are the annihilation (creation) operators for the flying and cavity modes, respectively. The parameter $\theta$ describes the finite cavity mirror transmittance $T = \cos^2 \theta$. We assume that each cavity mode is prepared in the vacuum state $|0\rangle_{c,\alpha}$ so that the initial state for the whole system is

$$\hat{\rho}_{cf}^{in}(\vec{x}) = \hat{\rho}_f(\vec{x}) \otimes |0\rangle_{c,A}\langle 0| \otimes |0\rangle_{c,B}\langle 0| \otimes |0\rangle_{c,C}\langle 0|. \quad (3)$$

After the interaction with the CV field the whole system density operator is given by:

$$\hat{\rho}_{cf}(\theta, \vec{x}) = \hat{B}_A(\theta)\hat{B}_B(\theta)\hat{B}_C(\theta)\hat{\rho}_{cf}^{in}(\vec{x})\hat{B}_A^\dagger(\theta)\hat{B}_B^\dagger(\theta)\hat{B}_C^\dagger(\theta). \quad (4)$$

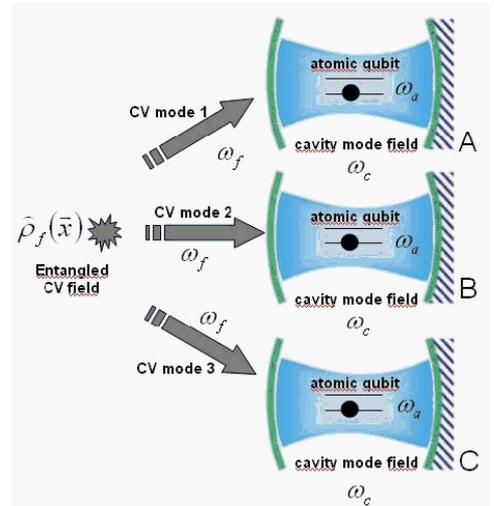

FIG. 1: Schematic diagram of the entanglement transfer process from a three-mode CV field to a system of three spatially separated two-level atoms, each one trapped inside a cavity.

Upon tracing out the field variables we obtain the density operator describing the state of the three cavity modes $\hat{\rho}_c(0, \theta, \vec{x}) = \text{Tr}_f [\hat{\rho}_{cf}(\theta, \vec{x})]$. The corresponding density matrix elements $c_{i,j,k,i',j',k'}(0, x, \theta) = \langle i, j, k|\hat{\rho}_c(0, \theta, \vec{x})|i', j', k'\rangle$ are related to those of the injected CV field by

$$c_{i,j,k,i',j',k'}(0, \theta, \vec{x}) = (\cos \theta)^{i+j+k+i'+j'+k'} \times$$
$$\times \sum_{l,m,n=0}^{\infty} a_{i+l,j+m,k+n,i'+l,j'+m,k'+n}(\vec{x})(\sin \theta)^{2(l+m+n)} \times$$
$$\times \Big[\frac{(i+l)!}{i!l!}\frac{(j+m)!}{j!m!}\frac{(k+n)!}{k!n!}\frac{(i'+l)!}{i'!l!}\frac{(j'+m)!}{j'!m!}\frac{(k'+n)!}{k'!n!}\Big]^{1/2}. \quad (5)$$

For unit transmittance $T = 1$ the operators $\hat{B}_\alpha(\theta)$ reduce to the identity and the injected CV field is fully transferred to the system of three cavities. Notice that within the beam splitter approach we assume that a) the atoms are injected into the cavities after the CV field has been transferred, and b) the atomic interaction time is shorter than the cavity mode lifetime. As discussed in [19] a more realistic approach may be used, upon taking into account cavity mode dissipation and atomic levels decay. Here we focus on the effectiveness of the transfer process for tripartite entanglement and use the beam splitter approach to simplify the treatment. The robustness of the scheme against dissipation and noise will be discussed elsewhere.

For the moment we assume that each atom is prepared in the ground state $|g\rangle_\alpha$; the initial density operator is thus given by

$$\hat{\rho}_{ac}(0, \theta, \vec{x}) = \rho_c(0, \theta, \vec{x}) \otimes |g\rangle_A\langle g| \otimes |g\rangle_B\langle g| \otimes |g\rangle_C\langle g| \quad (6)$$

The effect of different atomic preparations will be discussed in Sec. V. We also assume that each atom in-teracts resonantly with the cavity mode for a time $\tau$ shorter than the cavity decay time so that we can describe the interaction by the JC unitary operators $\hat{U}_\alpha(\tau)$ [42]. The atom-cavity density operator after the interaction is given by:

$$\hat{\rho}_{a,c}(\tau,\theta,\vec{x}) = \\ = \hat{U}_A(\tau)\hat{U}_B(\tau)\hat{U}_C(\tau)\rho_{ac}(0,\theta,\vec{x})\hat{U}_A^\dagger(\tau)\hat{U}_B^\dagger(\tau)\hat{U}_C^\dagger(\tau). \quad (7)$$

Finally, by taking the partial trace over the cavity mode variables we obtain the atomic density operator $\hat{\rho}_a(\tau,\theta,\vec{x}) = \text{Tr}_c\left[\hat{\rho}_{ac}(\tau,\theta,\vec{x})\right]$, whose corresponding density matrix elements in the standard basis:

$$\begin{aligned}\{|v_j\rangle_a\}_{j=1}^8 &= \{|e\rangle_A|e\rangle_B|e\rangle_C, |e\rangle_A|e\rangle_B|g\rangle_C, |e\rangle_A|g\rangle_B|e\rangle_C, \\ &\quad |e\rangle_A|g\rangle_B|g\rangle_C, |g\rangle_A|e\rangle_B|e\rangle_C, |g\rangle_A|e\rangle_B|g\rangle_C, \\ &\quad |g\rangle_A|g\rangle_B|e\rangle_C, |g\rangle_A|g\rangle_B|g\rangle_C\} \quad (8)\end{aligned}$$

are reported in the Appendix A, and $|e\rangle_\alpha$ ($|g\rangle_\alpha$) ($\alpha = \{A,B,C\}$) is the excited (ground) atomic state.

### A. Tripartite entanglement measure

In order to assess the effectiveness of the entanglement transfer process we need to check and possibly quantify the entanglement properties for the system of the three localized qubits. As far as we have to consider mixed states, the complete classification of three-qubit entanglement is still an open problem. In this paper, in order to measure the three-qubit entanglement, we employ the tripartite negativity $N_{ABC}$ introduced in [43], which is defined as:

$$N_{ABC} = \sqrt[3]{N_{A-BC} N_{B-AC} N_{C-AB}}, \quad (9)$$

It is the geometric mean of the negativities $N_{I-JK}$, with $I = A, B, C$ and $JK = BC, AC, AB$ which, in turn, are defined as $N_{I-JK} = -2\sum_i \sigma_i(\rho_a^{tI})$, where $\sigma_i(\rho_a^{tI})$ are the negative eigenvalues of the partial transpose $\rho_a^{tI}$ of the atomic density matrix with respect to the subsystem $I$.

Tripartite negativity improves the classification of three-qubit entanglement for pure states. For mixed states the positivity of $N_{ABC}$ excludes full separability or simple biseparability but cannot completely solve the problem of classifying full tripartite entanglement. In the following, we also consider the degree of mixedness of the atomic density operator, as measured by the purity $\mu_a = \text{Tr}_a\left[\hat{\rho}_a^2(\tau,\theta,\vec{x})\right]$, and evaluate the entanglement properties of the two-qubit subsystems, described by the reduced density operators obtained by tracing out one of the three qubits $\hat{\rho}_a^{(JK)} = \text{Tr}_I[\hat{\rho}_a]$ ($I = A,B,C$ and $JK = BC, AC, AB$).

## III. THREE-QUBIT-LIKE CV FIELDS

In this section we consider the case of CV fields in Eq. (1) such that coefficients $a_{p,q,r,p',q',r'}(\vec{x})$ are non vanishing only if all indexes are restricted to the values $\{0,1\}$, i.e. the CV state can be well approximated by a three-qubit state. This allows us to derive interesting analytical results for the entanglement transfer process and to describe situations where the nonlinearities used to generate tripartite entanglement in realistic CV fields are small.

### A. Pure states

For the state of three qubits it is possible to write different generalized Schmidt decompositions (GSD) using only five elements of the whole Hilbert space basis. In this section we consider the following form [43]:

$$|\psi_{GSD}\rangle_f = \alpha|000\rangle_f + \beta|100\rangle_f + \delta|110\rangle_f + \epsilon|101\rangle + \omega|111\rangle_f \quad (10)$$

that is symmetric in the interchange of the last two qubits. Any state of the three qubits can by reduced to $|\psi_{GSD}\rangle_f$ by suitable local unitary operations. We remark that totally symmetric GSD forms can be used as well [44], [45] and we briefly discuss them later. Here we focus on the $|\psi_{GSD}\rangle_f$ states because they can approximate an experimentally feasible CV field as we discuss in Sec. IV. We first investigate the case of perfect cavity mirror transmittance and we discuss different values of parameters in $|\psi_{GSD}\rangle_f$ showing that fully entangled tripartite atomic states can be obtained. Then we will evaluate the effect of mirror transmittance in a simple case.

#### 1. Perfect cavity mirror transmittance (T = 1)

In the case of perfect mirror transmittance the CV field is fully transferred to the cavity modes. In fact, if $T = 1$ the coefficients $c_{i,j,k,i',j',k'}(0,\theta,\vec{x})$ in Eq. (5) reduce to those of the injected CV field $a_{i,j,k,i',j',k'}(\vec{x})$. As discussed in [19] for the case of two-qubits systems this is an ideal case but it allows maximum entanglement transfer from driving fields to the atoms inside the cavities.
It is too complex to derive analytical expressions for the purity and the tripartite negativity of the atomic system if all the complex coefficients $\alpha, \beta, \delta, \epsilon, \omega$ in $|\psi_{GSD}\rangle_f$ assume non-zero values. Nevertheless, the problem can be solved numerically as we show in Fig. 2 in the case of real coefficients all equal to $\frac{1}{\sqrt{5}}$. From the atomic purity $\mu_a^{(GSD)}$ we see (Fig. 2a) that it is possible to obtain pure states of the three-qubits if the values of dimensionless interaction time $g\tau$ are multiples of $\frac{\pi}{2}$. The tripartite negativity $N_{ABC}^{(GSD)}$ also shows oscillations (Fig. 2b) but with a double period with respect to the purity; the max-ima of entanglement transfer ($N_{ABC}^{(GSD)} \cong 0.6$) occur only for odd multiples of $\frac{\pi}{2}$ (i.e. for dimensionless interaction times $g\tau_k = (2k+1)\frac{\pi}{2}$ with $k = 0, 1, 2....$). The periodicity in the purity and tripartite negativity functions can be found also for the other choices of coefficients in the state $|\psi_{GSD}\rangle_f$ and it is a typical effect of all atoms prepared in the ground state.

In order to understand the different periodicity of purity and tripartite negativity one should take into account that for even multiples of $\frac{\pi}{2}$ the atomic state, derived from Eqs. (A1) and (A2), is simply $\hat{\rho}_a = |u_8\rangle_a\langle u_8|$, that is the initial atomic state. On the other hand, for interaction times $g\tau_k$ the atomic pure states are:

$$|\psi\rangle_a^{(GSD)} = \mp i\alpha|v_8\rangle_a - \beta|v_4\rangle_a \pm i\delta|v_2\rangle_a \pm i\epsilon|v_3\rangle + \omega|v_1\rangle_a \quad (11)$$

where the upper(lower) sign stands for even (odd) values of $k$. We note that the state $|\psi\rangle_a^{(GSD)}$ is closely related to the form of the injected CV state in Eq. (10). Following the classification proposed in [43] the pure atomic state in Eq. (11) can be separable, bi-separable or fully entangled depending on the number of coefficients set to zero.

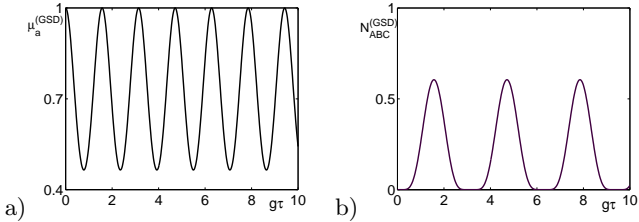

FIG. 2: Entanglement transfer for a field state $|\psi_{GSD}\rangle_f$ with all coefficients equal to $\frac{1}{\sqrt{5}}$, as a function of interaction time $g\tau$. a) Atomic purity $\mu_a^{(GSD)}$, b) tripartite negativity $N_{ABC}^{(GSD)}$.

In the case $\delta = 0$ and all the other coefficients non-zero, the pure atomic state in Eq. (11) is of subtype 2-2 (star-shaped). For the partial transpose matrices $\rho_a^{tB}$ and $\rho_a^{tA}$ we obtain the same eigenvalues that are different from those of $\rho_a^{tC}$, and for the tripartite negativity we get:

$$N_{ABC}^{(2-2)} = 2\{|\alpha||\omega|\sqrt{|\alpha|^2 + |\beta|^2} \times \\ \times \sqrt{|\epsilon|^2 + |\omega|^2}\sqrt{|\beta|^2|\omega|^2 + |\alpha|^2(|\epsilon|^2 + |\omega|^2)}\}^{\frac{1}{3}} \quad (12)$$

If we consider the two-qubit subsystems we find that the state $\hat{\rho}_a^{(AB)}$ is fully separable while $\hat{\rho}_a^{(BC)}$, $\hat{\rho}_a^{(AC)}$ states are entangled.

In the case $\delta = \epsilon = 0$ we change to an atomic state of the subtype 2-1 with tripartite negativity $N_{ABC}^{(2-1)} = 2|\omega|\sqrt[3]{|\alpha|(1-|\omega|^2)}$. For the two-qubit subsystems we have two unentangled states $\hat{\rho}_a^{(AB)}$ and $\hat{\rho}_a^{(AC)}$ while $\hat{\rho}_a^{(BC)}$ is entangled with negativity $N_{BC} = 2|\beta||\omega|$.

If we set the parameters $\delta = \epsilon = \beta = 0$ the pure atomic state changes to subtype 2-0 (GHZ-like states). The tripartite negativity reduces to $N_{ABC}^{(2-0)} = 2|\alpha||\omega|$ and for the 2-qubit subsystems we find three identical unentangled mixed states.

To obtain atomic states in Eq. (11) corresponding to the subtype 2-3 (W-like states) described in [43], we set $\beta = \omega = 0$ and all the other coefficients are non-zero. The tripartite negativity is:

$$N_{ABC}^{(2-3)} = 2\sqrt[3]{|\alpha||\delta||\epsilon|}\sqrt{1-|\alpha|^2}\sqrt{1-|\delta|^2}\sqrt{1-|\epsilon|^2} \quad (13)$$

and for the two-qubit subsystems we find three entangled states.

Finally, we evaluate the effect of interaction time out of $g\tau_k$ values for $|\psi_{GSD}\rangle_f$ in the case $\gamma = \beta = 0$ (subtype 2-3) because this kind of CV field can be experimentally realized as discussed in Sec. IV. The atomic purity $\mu_a^{(2-3)}(g\tau)$ can be written as a function of only the coefficient $|\alpha|^2$ as:

$$\mu_a^{(2-3)}(g\tau) = \frac{1}{16}\left\{|\alpha|^2 + (1-|\alpha|^2)[3 + \cos(4g\tau)]\right\}^2 \quad (14)$$

In Fig. 3a we show $\mu_a^{(2-3)}(g\tau)$ and we see that for fixed $g\tau_k$ values it is an increasing function of $|\alpha|^2$. In Fig. 3b we show the tripartite negativity $N_{ABC}^{(2-3)}$ evaluated numerically as a function of $|\alpha|^2$ and $g\tau$ in the case of $|\delta|^2 = |\epsilon|^2 = \frac{1-|\alpha|^2}{2}$. Again we find that the entanglement transfer is more effective for interaction times $g\tau_k$ and for $|\alpha|^2 = |\delta|^2 = |\epsilon|^2 = \frac{1}{3}$ we obtain the maximum of tripartite negativity ($\frac{2\sqrt{2}}{3}$). In addition, in Figs. 3c,d we show the negativities $N_{JK}$ of two-qubit subsystems.

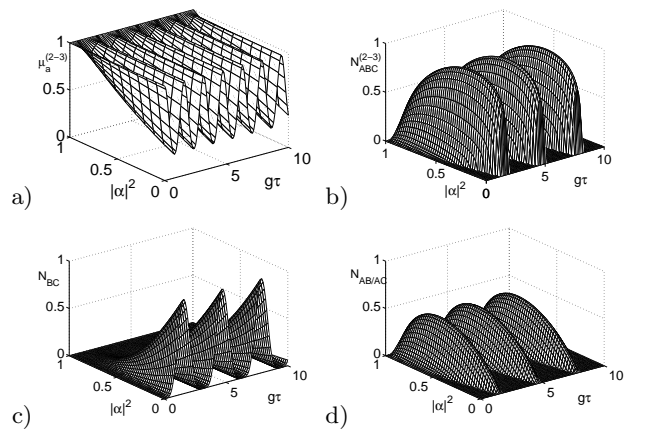

FIG. 3: Entanglement transfer for a field state $|\psi_{GSD}\rangle_f$ with $\beta = \omega = 0$ vs. $g\tau$ and $|\alpha|^2$. a) Atomic purity $\mu_a^{(2-3)}$ as in Eq. (14), b) tripartite negativity $N_{ABC}^{(2-3)}$ evaluated numerically for $|\delta|^2 = |\epsilon|^2$, c-d) negativity of two-qubit subsystems $N_{BC}$ and $N_{AB} = N_{AC}$.

2. *The effect of losses for mirror transmittance $T < 1$*

To evaluate the effect of cavity mirror transmittance $T < 1$ on the entanglement transfer process we consider

$|\psi_{GSD}\rangle_f$ in Eq. (10) with $\delta = \beta = \epsilon = 0$ (GHZ-like states). In this case the non vanishing atomic density matrix elements can be written as:

$$\begin{aligned}
\rho_{a,11} &= |\omega|^2 T^3 \sin^6(g\tau) \\
\rho_{a,22} &= \rho_{a,33} = \rho_{a,55} = |\omega|^2 T^2 \sin^4(g\tau) \\
\rho_{a,44} &= \rho_{a,66} = \rho_{a,77} = |\omega|^2 T \sin^2(g\tau)[1 - T\sin^2(g\tau)]^2 \\
\rho_{a,88} &= |\alpha|^2 + |\omega|^2[1 - T\sin^2(g\tau)]^2 \\
\rho_{a,18} &= i\omega\alpha^* T^{\frac{3}{2}} \sin^3(g\tau)
\end{aligned} \quad (15)$$

For the atomic purity $\mu_a^{(2-0)}(g\tau, T)$ we derive:

$$\mu_a^{(2-0)}(g\tau, T) = |\alpha|^4 + |\omega|^4 \Big[1 - 2Y(1-Y)\Big]^3 + \\ + 2|\omega|^2|\alpha|^2 \Big[1 - 3Y(1-Y)\Big] \quad (16)$$

where we introduced $Y \equiv T\sin(g\tau)^2$ to simplify the notations. All partial transpose matrices $\rho_a^{tI}$ have the same eigenvalue $\lambda^-$ that may assume negative values:

$$\begin{aligned}
\lambda^- &= \frac{1}{2}|\omega|Y\bigg\{|\omega|(1-T) - \Big[4|\alpha|^2 Y + \\
&+ |\omega|^2\Big(1 - 6Y + Y^2(13 - 12Y + 4Y^2)\Big)\Big]^{\frac{1}{2}}\bigg\}
\end{aligned} \quad (17)$$

In Fig. 4 we consider the case of GHZ state ($|\alpha|^2 = |\omega|^2 = \frac{1}{2}$) because it allows maximum entanglement transfer for $T = 1$. We see that the main effect of decreasing $T$ is a progressive reduction of the atomic tripartite negativity but for $T > 0.5$ we can transfer entanglement significantly.

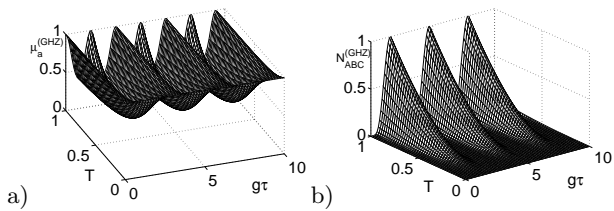

FIG. 4: The effect of cavity mirror transmittance $T$ in the case of $|\psi_{GSD}\rangle_f$ in a GHZ state ($|\alpha|^2 = |\omega|^2 = \frac{1}{2}$). a) Atomic purity $\mu_a^{(GHZ)}(g\tau, T)$ as in Eq. (16), b) tripartite negativity $N_{ABC}^{(GHZ)}(g\tau, T)$ numerically evaluated from Eq. (17).

### 3. Fully symmetric GSD states

In this section we discuss the entanglement transfer in the cases of CV fields approximated by states of three-qubits in GSD forms that are symmetric with respect to any exchange of the qubit pairs and we compare them with the above case of states $|\psi_{(GSD)}\rangle_f$ in Eq. (10). In particular, we focus on the following two GSD forms ([44], [45]):

$$\begin{aligned}
|\phi_{3s}\rangle_f &= a|000\rangle_f + b|001\rangle_f + c|010\rangle_f + d|100\rangle_f + e|111\rangle_f \\
|\varphi_{3s}\rangle_f &= a|000\rangle_f + b|011\rangle_f + c|101\rangle_f + d|110\rangle_f + e|111\rangle_f
\end{aligned} \quad (18)$$

From the numerical results shown in Fig. 5 we see that the GSD form $|\phi_{3s}\rangle_f$ seems to be more efficient and almost optimal for the entanglement transfer, also generating atomic states with a higher degree of purity.

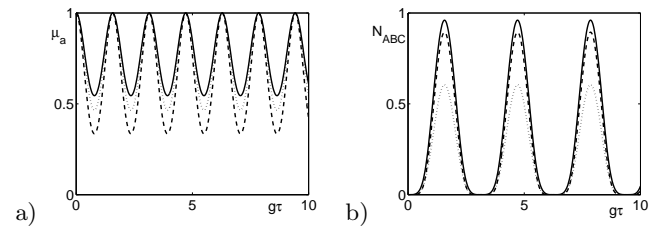

FIG. 5: Entanglement transfer for different GSD forms of CV fields approximated by three qubit states. We consider all the coefficients of the GSD states equal to $\frac{1}{\sqrt{5}}$. a) Atomic purity $\mu_a$, b) tripartite negativity $N_{ABC}$. We consider the states $|\psi_{(GSD)}\rangle_f$ (dot), $|\varphi_{3s}\rangle_f$ (dashed), $|\phi_{3s}\rangle_f$ (solid)

### B. Mixed state

As an example of CV fields prepared in a mixed state we consider [46]:

$$\hat{\rho}_f^{(M)}(p) = p|GHZ\rangle_f\langle GHZ| + (1-p)|W'\rangle_f\langle W'| \quad (19)$$

where $p$ is a real parameter, $|GHZ\rangle_f = \frac{1}{\sqrt{2}}(|000\rangle_f + |111\rangle_f)$, and $|W'\rangle_f = \frac{1}{\sqrt{3}}(|001\rangle_f + |010\rangle_f + |100\rangle_f)$. We note that state $|W'\rangle_f$ is the original W-state written in the case of a fully symmetric GSD decomposition $|\phi_{3s}\rangle_f$ in Eq. (18). In the case of cavity mirror transmittance $T = 1$ we can derive the following non vanishing density matrix elements of the three-qubit system:

$$\begin{aligned}
\rho_{a,11} &= \frac{p}{2}\sin^6(g\tau) \\
\rho_{a,22} &= \rho_{a,33} = \rho_{a,55} = \frac{p}{2}\sin^4(g\tau)\cos^2(g\tau) \\
\rho_{a,44} &= \rho_{a,66} = \rho_{a,77} = \sin^2(g\tau)\left[\frac{1-p}{3} + \frac{p}{2}\cos^4(g\tau)\right] \\
\rho_{a,88} &= \frac{p}{2}(1 + \cos^6(g\tau)) + (1-p)\cos^2(g\tau) \\
\rho_{a,18} &= i\frac{p}{2}\sin^3(g\tau) \\
\rho_{a,46} &= \rho_{a,47} = \rho_{a,67} = \frac{1-p}{3}\sin^2(g\tau)
\end{aligned} \quad (20)$$

We report in the Appendix B the expressions derived for the atomic purity $\mu_a^{(M)}(g\tau, p)$ (Fig. 6a) and for the eigenvalues of the partial transpose matrices $\rho_a^{tI}$ from which we can numerically evaluate the tripartite negativity $N_{ABC}^{(M)}(g\tau, p)$ (Fig. 6b). We see large entanglement transfer for dimensionless interaction times $g\tau_k = (2k+1)\frac{\pi}{2}$ ($k = 0, 1, 2, ...$). The maxima of tripartite negativity can be found in the limits of CV field with $p = 0$ (pure $W'$-state) and $p = 1$ (pure GHZ-state) and are equal to $\frac{2\sqrt{2}}{3}$ and 1, respectively. We remark that for interaction times $g\tau_k$ the atomic state reduces to the form of Eq. (19) but with the GHZ term of the form $|GHZ_\pm\rangle_a = \frac{1}{\sqrt{2}}(|000\rangle_a \pm i|111\rangle_a)$, where $+(-)$ holds for even (odd) values of the integer $k$. Therefore, if we consider the limit $p = 0$ we find that the CV state $|W'\rangle_f$ is exactly transferred to the atomic system. We find that the atomic purity is simply $\mu_a^{(M)}(g\tau_k, p) = 2p^2 - 2p + 1$ and the tripartite negativity

$$N_{ABC}^{(M)}(g\tau_k, p) =$$
$$= \frac{2\sqrt{10p^2 - 2p + 1} + \sqrt{41p^2 - 64p + 32} - p - 2}{6}$$
(21)

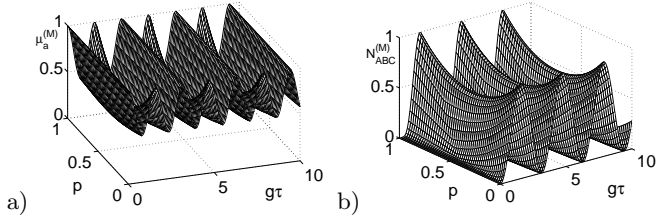

FIG. 6: Entanglement transfer from a mixed CV field as in Eq. (19) as a function of interaction time $g\tau$ and parameter $p$. a) Atomic purity $\mu_a^{(M)}$, b) tripartite negativity $N_{ABC}^{(M)}$.

## IV. PURE GAUSSIAN CV FIELDS

In this section we consider as an example of CV field the experimentally feasible state proposed in [38]. It is a Gaussian fully inseparable three-mode radiation generated by two type-I noncollinearly phase-matched interlinked bilinear interactions that simultaneously couple the three modes. This state was proposed to realize a telecloning protocol in noisy channel [47].
The generation process of the above state can be described by the following interaction Hamiltonian:

$$H_{int} = \tilde{\gamma}_1 \hat{a}_1^\dagger \hat{a}_3^\dagger + \tilde{\gamma}_2 \hat{a}_2^\dagger \hat{a}_3 + h.c \quad (22)$$

The effective coupling constant $\tilde{\gamma}_j$ ($j = 1, 2$) of the two parametric processes are proportional to the nonlinear susceptibilities and the pump intensities. In the three mode Fock basis $|pqr\rangle_f$ the outgoing state after an interaction time $\tau_f$ is given by:

$$|T\rangle_f = \frac{1}{\sqrt{1 + N_1}} \sum_{p,q=0}^{\infty} \left[\frac{N_2}{1 + N_1}\right]^{\frac{p}{2}} \times$$
$$\times \left[\frac{N_3}{1 + N_1}\right]^{\frac{q}{2}} \left[\frac{(p+q)!}{p!q!}\right]^{\frac{1}{2}} |p+q, p, q\rangle_f \quad (23)$$

where $N_i = \langle \hat{a}_j^\dagger \hat{a}_j \rangle$ ($i = 1, 2, 3$) is the average number of photons in the i-th mode. We have $N_1 = N_2 + N_3$ and the mean photon numbers $N_i$ ($i = 2, 3$) are related to the dimensionless coupling constants $\gamma_j = \tau_f \tilde{\gamma}_j$ by:

$$N_2 = \frac{|\gamma_1|^2 |\gamma_2|^2}{(|\gamma_2|^2 - |\gamma_1|^2)^2} \left[1 - \cos(\sqrt{|\gamma_2|^2 - |\gamma_1|^2})\right]^2$$
$$N_3 = \frac{|\gamma_1|^2}{|\gamma_2|^2 - |\gamma_1|^2} \sin^2(\sqrt{|\gamma_2|^2 - |\gamma_1|^2}) \quad (24)$$

We recall that the above Hamiltonian was first considered in [48] and more recently in the treatment of the collective atomic recoil laser [49].
First we consider the case of equal coupling constants ($\gamma_1 = \gamma_2$) so that the field mode mean photon numbers are simply given by $N_2 = \frac{|\gamma_1|^4}{4}$ and $N_3 = |\gamma_1|^2$. We numerically evaluate the atomic density matrix elements derived in Appendix A in the case of mirror transmittance $T = 1$. In Fig. 7a we show the atomic purity $\mu_a^{(T)}$ as function of the CV field parameter $|\gamma_1|^2$ and the dimensionless interaction time $g\tau$ between the atoms and the cavity mode fields. We see that the atomic purity decreases for increasing values of coupling parameters for any interaction time. In Fig. 7b we show the tripartite negativity $N_{ABC}^{(T)}$ and we see that regions with large entanglement transfer can be found around the interaction times $g\tau_k = (2k+1)\frac{\pi}{2}$ ($k = 0, 1, 2, ...$). The maxima of tripartite negativity occur for a coupling parameter value $|\gamma_1|^2 \cong 0.6$. Increasing $|\gamma_1|^2$ we increase the mean photon numbers in the CV field modes and we note a reduction in the tripartite negativity that can be explained as follows. The photonstatistics of state $|T\rangle_f$ contains only three non-zero coefficients such that $n, m, s = \{0, 1\}$ whose probabilities are given by $|b_{110}|^2 = N_2(1 + N_2 + N_3)^{-2}$, $|b_{101}|^2 = N_3(1+N_2+N_3)^{-2}$ and $|b_{000}|^2 = (1+N_2+N_3)^{-1}$. In Fig. 7c we show these probabilities as functions of the CV field coupling parameter and we see that for $|\gamma_1|^2 < 1$ the photonstatistics is nearly saturated by the above terms. Therefore, the $|T\rangle_f$ state can be in fact well approximated by a W-like state (see subsection III A 1). For larger values of $|\gamma_1|^2$ the photonstatistics contains an increasing number of terms, that is a less favorable condition for entanglement transfer between the CV field and the atomic system. We remark that the value $|\gamma_1|^2 = 0.6$ of the maxima of tripartite negativity corresponds to the maximum of probability $|b_{101}|^2$.

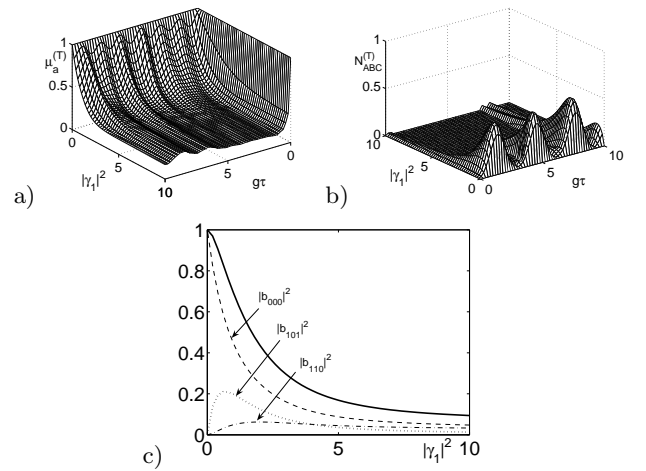

FIG. 7: Entanglement transfer for a CV field in the state of Eq.(23) with equal coupling parameters $\gamma_j$. a) Atomic purity $\mu_a^{(T)}(g\tau, |\gamma_1|^2)$, b) tripartite negativity $N_{ABC}^{(T)}(g\tau, |\gamma_1|^2)$, c) the first terms of $|T\rangle_f$ state photonstatistics $|b_{000}|^2$ (dashed), $|b_{101}|^2$ (dotted), $|b_{110}|^2$(dash-dotted), and their sum (solid line).

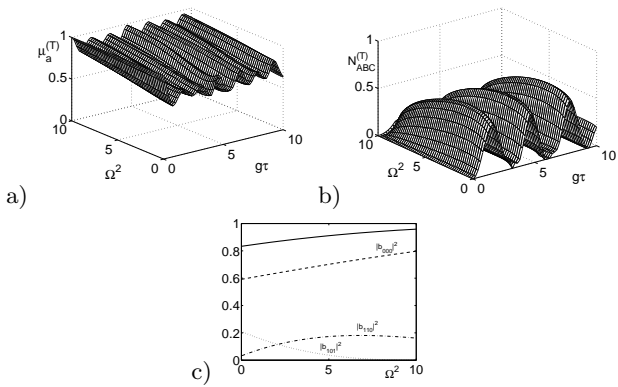

FIG. 8: Entanglement transfer from a CV field in the state of Eq. (23) with $|\gamma_1|^2 = 0.6$. a) Atomic purity $\mu_a^{(T)}(g\tau, \Omega^2)$, b) tripartite negativity $N_{ABC}^{(T)}(g\tau, \Omega^2)$, c) the first terms of $|T\rangle$ state photonstatistics $|b_{000}|^2$ (dashed), $|b_{101}|^2$ (dotted), $|b_{110}|^2$ (dash-dotted), and their sum (solid line).

In order to describe the effect on the entanglement transfer process of different coupling constants $\gamma_1$ and $\gamma_2$ we introduce the parameter $\Omega = \sqrt{|\gamma_2|^2 - |\gamma_1|^2}$. As an example, in Fig. 8 we consider the case of $|\gamma_1|^2 = 0.6$ because it corresponds to the maxima of tripartite negativity in Fig. 7b. In Fig. 8a we see that the atomic purity $\mu_a^{(T)}$ shows oscillations as a function of $g\tau$ and it is a slowly varying function of parameter $\Omega^2$. In Fig. 8b we see that for interaction times $g\tau_k = (2k+1)\frac{\pi}{2}$ ($k = 0, 1, 2, ...$) large entanglement transfer is possible and the tripartite negativity $N_{ABC}^{(T)}$ is again a slowly varying function of parameter $\Omega^2$ because the CV field photonstatistics (Fig. 8c) is nearly saturated by the first terms $|b_{000}|^2$, $|b_{110}|^2$ and $|b_{101}|^2$ in the whole range of $\Omega^2$. In addition, we note that the maxima of tripartite negativity occur for $\Omega^2 \cong 5$, corresponding to the maximum of $|b_{110}|^2$. Upon assuming a fixed interaction time $g\tau_k = \frac{\pi}{2}$ we may investigate the effect of both CV field coupling parameters on the entanglement transfer. In particular, in Fig. 9b we see that the behavior of the tripartite negativity $N_{ABC}^{(T)}$ can be distinguished into two main regions corresponding to small and large values of the mean photon numbers $N_J$ (Fig. 9c,d) confirming that large values of $N_2$ and $N_3$ do not favorite the entanglement transfer.

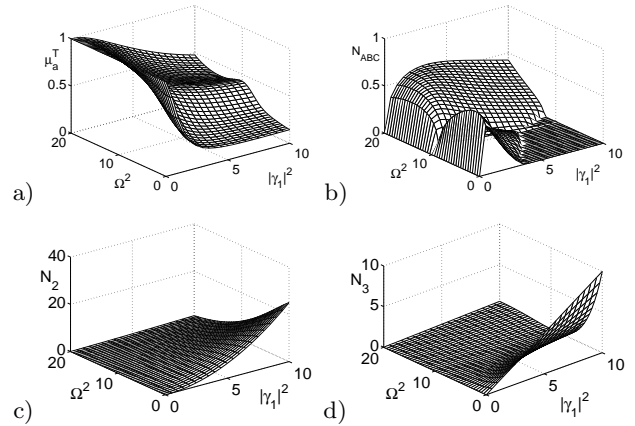

FIG. 9: Entanglement transfer from a CV field in the state of Eq. (23) as a function of parameters $|\gamma_1|^2$ and $\Omega^2$ for fixed atom-field interaction time $g\tau = \frac{\pi}{2}$. a) Atomic purity $\mu_a^{(T)}$, b) tripartite negativity $N_{ABC}^{(T)}$, c-d) mean photon numbers $N_2$ and $N_3$.

Finally, it is interesting to evaluate the entanglement properties of the bipartite subsystems described by the reduced density operators $\hat{\rho}_a^{(JK)}$ with $JK = BC, AC, AB$. We find that $\hat{\rho}_a^{(BC)}$ is not entangled for any value of the CV field parameters while in Figs. 10a,b we see regions where the negativities $N_{AB}$ and $N_{AC}$ are not vanishing. This reflects the peculiarity of $|T\rangle_f$ states [49], where the two couples of modes $1 - 2$ and $1 - 3$ are entangled subsystems while the subsystem of modes $2 - 3$ is not entangled for any choice of CV field coupling constants.

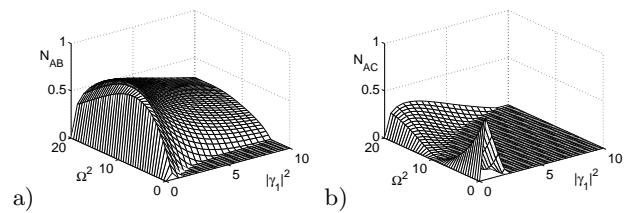

FIG. 10: Entanglement transfer from a CV field in the state of Eq.(23): entanglement properties of atomic bipartite subsystems $\hat{\rho}_a^{(JK)}$ as functions of parameters $|\gamma_1|^2$ and $\Omega^2$ for fixed interaction time $g\tau = \frac{\pi}{2}$. a) Negativity $N_{AB}$, b) negativity $N_{AC}$. The negativity of subsystem $BC$ is always zero.

## V. THE EFFECT OF ATOMIC PREPARATIONS

In the above sections we investigated the entanglement transfer in the case of all atoms prepared in the ground state. Here we briefly discuss the effect of preparing the atoms in different separable initial states. Depending on the choice of the initial atomic state we can derive specific expressions for the atomic density matrix elements in the standard basis like those listed in Eqs. (A1),(A2). We only remark that the trigonometric functions contain terms like $\sqrt{2}g\tau$ and not only $g\tau$ as in the case of the atomic state $|ggg\rangle_a$ discussed above. This explains the more complex dependence on interaction time as those illustrated in the following examples. We first consider the CV field $|\psi_{GSD}\rangle_f$ in Eq. (10) and we focus on the case of all coefficients equal to $\frac{1}{\sqrt{5}}$ to compare the results with those in Fig. 2. In Fig. 11a we see in general that an atomic preparation different from $|ggg\rangle_a$ state implies a higher degree of mixedness in the final atomic state. Exceptions can be found for dimensionless interaction times $g\tau_k = \frac{3}{2}\pi$ and $\frac{5}{2}\pi$ but Fig. 11b shows that the tripartite negativity is less affected by the change of the atomic preparation if $g\tau_k = \frac{3\pi}{2}$.

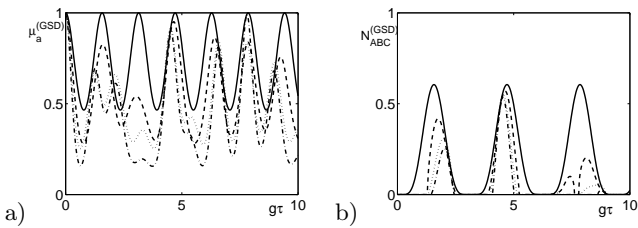

FIG. 11: Effect of different atomic preparation in the entanglement transfer for $|\psi_{GSD}\rangle_f$ states in the case of all coefficients equal to $\frac{1}{\sqrt{5}}$. a) Atomic purity $\mu_a^{(GSD)}$, b) tripartite negativity $N_{ABC}^{(GSD)}$. We consider the initial atomic states $|ggg\rangle_a$ (solid), $|geg\rangle_a$ (dash), $|ege\rangle_a$ (dot), and $|eee\rangle_a$ (dash-dot).

We first consider the CV field $|\psi_{GSD}\rangle_f$ in Eq. (10) and we focus on the case of all coefficients equal to $\frac{1}{\sqrt{5}}$ to compare the results with those in Fig. 2. In Fig. 11a we see in general that an atomic preparation different from $|ggg\rangle_a$ state implies a higher degree of mixedness in the final atomic state. Exceptions can be found for dimensionless interaction times $g\tau_k = \frac{3}{2}\pi$ and $\frac{5}{2}\pi$ but Fig. 11b shows that the tripartite negativity is less affected by the change of the atomic preparation if $g\tau_k = \frac{3\pi}{2}$.

Finally, we consider the CV field prepared in the $|T\rangle_f$ state and limit ourselves to the case of fixed parameters $\Omega = 0$ and $|\gamma_1|^2 = 0.6$. In Fig. 12 we show the dependence of the atomic purity and the tripartite negativity on the dimensionless interaction time $g\tau$. Again the choice of preparing all atoms in the ground state leads to a larger purity and entanglement of the atomic state at the end of the process.

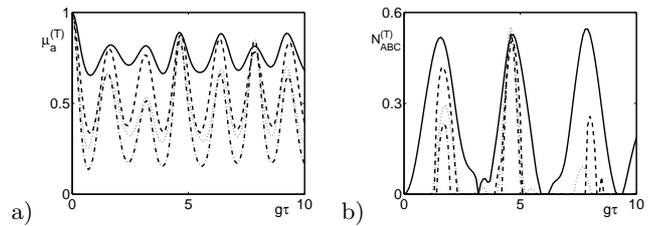

FIG. 12: Effect of different atomic preparation in the entanglement transfer for CV field in the $|T\rangle_f$ state of Eq. (23) with $\Omega^2 = 0$ and $|\gamma_1|^2 = 0.6$. a) Atomic purity $\mu_a^{(GSD)}$ and b) tripartite negativity $N_{ABC}^{(GSD)}$ for initial atomic states $|ggg\rangle_a$ (solid), $|geg\rangle_a$ (dash), $|ege\rangle_a$ (dot), and $|eee\rangle_a$ (dash-dot).

## VI. CONCLUSIONS

We have investigated in details the problem of entanglement transfer between a three-mode state of the field and a system of three separated qubits (i.e. two-level atoms) each one interacting with a local environment (i.e. a cavity mode). We have shown that large atomic entanglement may be obtained in the case of field states approximated by a three-qubit-like states for suitable dimensionless interaction times. We investigated a wide range of radiation states in the qubit-like form (pure and mixed) and derive analytical formulas for the tripartite negativity as well as for the purity of the three-qubit system and two-qubit subsystems. We have also compared different Schmidt decompositions for the qubit-like field states showing that the fully symmetric ones are the most efficient for entanglement transfer. The case of an experimentally feasible Gaussian three-mode entangled field state showing which range of field parameters should be selected to obtain a large entanglement transfer. In addition, we also discussed different atomic preparations showing that the case of all atoms prepared in the ground state allows a nice periodicity in the entanglement transfer as function of the interaction time.

The results of this paper extend and complete those obtained in our previous works [18, 19] about the entanglement transfer from two-mode states of the field to two-qubit systems. The extension to the case of tripartite systems is not trivial due to the the higher degree of complexity of the problem and also for the peculiar structure of tripartite entanglement. In our treatment we choose to describe the entanglement transfer process under the assumptions of fully resonant interactions and strong coupling regimes in order to avoid the effects of atomic decay and cavity mode dissipation through the environment. We are currently investigating the above effects by the Monte Carlo Wave Function techniques in the perspective of an experimental realization of our scheme for quantum memories and quantum computing.

**Acknowledgments**

This work has been supported by MIUR through the project PRIN-2005024254-002.

## APPENDIX A: ATOMIC DENSITY MATRIX ELEMENTS

In this Appendix we report the results for the entanglement transfer between the CV field in Eq. (1) and a system of three two-level atoms, all prepared in their ground state, and located in spatially separated cavities each one prepared in the vacuum state. In the three-mode Fock basis $|pqr\rangle_f = \{|p\rangle_1 \otimes |q\rangle_2 \otimes |r\rangle_3\}_{p,q,r=0}^{\infty}$ the CV field, that can be in general a mixed state, is described by the complex coefficients $a_{p,q,r,p',q',r'}(\vec{x})$. After the CV field-cavity mode resonant interactions we trace out the field variables and we derive the cavity mode state described by coefficients $a_{p,q,r,p',q',r'}(\vec{x})$ in Eq. (5). After the atom-cavity mode Jaynes-Cummings resonant interaction we trace out the cavity mode variables and we derive the following elements of the $8 \times 8$ atomic density matrix $\rho_a(\tau, \theta, \vec{x})$, written in the standard basis of Eq. (8). The diagonal matrix elements are given by:

$$\begin{aligned}
\rho_{a,11} &= \sum_{i,j,k=0}^{\infty} c_{i+1,j+1,k+1,i+1,j+1,k+1}(0,\theta,x) \sin^2(g\tau\sqrt{i+1}) \sin^2(g\tau\sqrt{j+1}) \sin^2(g\tau\sqrt{k+1}) \\
\rho_{a,22} &= \sum_{i,j,k=0}^{\infty} c_{i+1,j+1,k,i+1,j+1,k}(0,\theta,x) \sin^2(g\tau\sqrt{i+1}) \sin^2(g\tau\sqrt{j+1}) \cos^2(g\tau\sqrt{k}) \\
\rho_{a,33} &= \sum_{i,j,k=0}^{\infty} c_{i+1,j,k+1,i+1,j,k+1}(0,\theta,x) \sin^2(g\tau\sqrt{i+1}) \cos^2(g\tau\sqrt{j}) \sin^2(g\tau\sqrt{k+1}) \\
\rho_{a,44} &= \sum_{i,j,k=0}^{\infty} c_{i+1,j,k,i+1,j,k}(0,\theta,x) \sin^2(g\tau\sqrt{i+1}) \cos^2(g\tau\sqrt{j}) \cos^2(g\tau\sqrt{k}) \\
\rho_{a,55} &= \sum_{i,j,k=0}^{\infty} c_{i,j+1,k+1,i,j+1,k+1}(0,\theta,x) \cos^2(g\tau\sqrt{i}) \sin^2(g\tau\sqrt{j+1}) \sin^2(g\tau\sqrt{k+1}) \\
\rho_{a,66} &= \sum_{i,j,k=0}^{\infty} c_{i,j+1,k,i,j+1,k}(0,\theta,x) \cos^2(g\tau\sqrt{i}) \sin^2(g\tau\sqrt{j+1}) \cos^2(g\tau\sqrt{k}) \\
\rho_{a,77} &= \sum_{i,j,k=0}^{\infty} c_{i,j,k+1,i,j,k+1}(0,\theta,x) \cos^2(g\tau\sqrt{i}) \cos^2(g\tau\sqrt{j}) \sin^2(g\tau\sqrt{k+1}) \\
\rho_{a,88} &= \sum_{i,j,k=0}^{\infty} c_{i,j,k,i,j,k}(0,\theta,x) \cos^2(g\tau\sqrt{i}) \cos^2(g\tau\sqrt{j}) \cos^2(g\tau\sqrt{k})
\end{aligned}$$

(A1)

Recalling that $\rho_{a,ji} = \rho_{a,ij}^*$, the off-diagonal elements of atomic density matrix are:

$$\rho_{a,12} = -i \sum_{i,j,k=0}^{\infty} c_{i+1,j+1,k+1,i+1,j+1,k}(0,\theta,x) \sin^2(g\tau\sqrt{i+1}) \sin^2(g\tau\sqrt{j+1}) \sin(g\tau\sqrt{k+1}) \cos(g\tau\sqrt{k})$$

$$\rho_{a,13} = -i \sum_{i,j,k=0}^{\infty} c_{i+1,j+1,k+1,i+1,j,k+1}(0,\theta,x) \sin^2(g\tau\sqrt{i+1}) \sin(g\tau\sqrt{j+1}) \cos(g\tau\sqrt{j}) \sin^2(g\tau\sqrt{k+1})$$

$$\rho_{a,14} = - \sum_{i,j,k=0}^{\infty} c_{i+1,j+1,k+1,i+1,j,k}(0,\theta,x) \sin^2(g\tau\sqrt{i+1}) \sin(g\tau\sqrt{j+1}) \cos(g\tau\sqrt{j}) \sin(g\tau\sqrt{k+1}) \cos(g\tau\sqrt{k})$$

$$\rho_{a,15} = -i \sum_{i,j,k=0}^{\infty} c_{i+1,j+1,k+1,i,j+1,k+1}(0,\theta,x) \sin(g\tau\sqrt{i+1}) \cos(g\tau\sqrt{i}) \sin^2(g\tau\sqrt{j+1}) \sin^2(g\tau\sqrt{k+1})$$

$$\rho_{a,16} = - \sum_{i,j,k=0}^{\infty} c_{i+1,j+1,k+1,i,j+1,k}(0,\theta,x) \sin(g\tau\sqrt{i+1}) \cos(g\tau\sqrt{i}) \sin^2(g\tau\sqrt{j+1}) \sin(g\tau\sqrt{k+1}) \cos(g\tau\sqrt{k})$$

$$\rho_{a,17} = - \sum_{i,j,k=0}^{\infty} c_{i+1,j+1,k+1,i,j,k+1}(0,\theta,x) \sin(g\tau\sqrt{i+1}) \cos(g\tau\sqrt{i}) \sin(g\tau\sqrt{j+1}) \cos(g\tau\sqrt{j}) \sin^2(g\tau\sqrt{k+1})$$

$$\rho_{a,18} = i \sum_{i,j,k=0}^{\infty} c_{i+1,j+1,k+1,i,j,k}(0,\theta,x) \sin(g\tau\sqrt{i+1}) \cos(g\tau\sqrt{i}) \sin(g\tau\sqrt{j+1}) \cos(g\tau\sqrt{j}) \sin(g\tau\sqrt{k+1}) \cos(g\tau\sqrt{k})$$

$$\rho_{a,23} = \sum_{i,j,k=0}^{\infty} c_{i+1,j+1,k,i+1,j,k+1}(0,\theta,x) \sin^2(g\tau\sqrt{i+1}) \sin(g\tau\sqrt{j+1}) \cos(g\tau\sqrt{j}) \sin(g\tau\sqrt{k+1}) \cos(g\tau\sqrt{k})$$

$$\rho_{a,24} = -i \sum_{i,j,k=0}^{\infty} c_{i+1,j+1,k,i+1,j,k}(0,\theta,x) \sin^2(g\tau\sqrt{i+1}) \sin(g\tau\sqrt{j+1}) \cos(g\tau\sqrt{j}) \cos^2(g\tau\sqrt{k})$$

$$\rho_{a,25} = \sum_{i,j,k=0}^{\infty} c_{i+1,j+1,k,i,j+1,k+1}(0,\theta,x) \sin(g\tau\sqrt{i+1}) \cos(g\tau\sqrt{i}) \sin^2(g\tau\sqrt{j+1}) \cos(g\tau\sqrt{k}) \sin(g\tau\sqrt{k+1})$$

$$\rho_{a,26} = -i \sum_{i,j,k=0}^{\infty} c_{i+1,j+1,k,i,j+1,k}(0,\theta,x) \sin(g\tau\sqrt{i+1}) \cos(g\tau\sqrt{i}) \sin^2(g\tau\sqrt{j+1}) \cos^2(g\tau\sqrt{k})$$

$$\rho_{a,27} = -i \sum_{i,j,k=0}^{\infty} c_{i+1,j+1,k,i,j,k+1}(0,\theta,x) \sin(g\tau\sqrt{i+1}) \cos(g\tau\sqrt{i}) \sin(g\tau\sqrt{j+1}) \cos(g\tau\sqrt{j}) \cos(g\tau\sqrt{k}) \sin(g\tau\sqrt{k+1})$$

$$\rho_{a,28} = -i \sum_{i,j,k=0}^{\infty} c_{i+1,j+1,k,i,j,k}(0,\theta,x) \sin(g\tau\sqrt{i+1}) \cos(g\tau\sqrt{i}) \sin(g\tau\sqrt{j+1}) \cos(g\tau\sqrt{j}) \cos^2(g\tau\sqrt{k})$$

$$\rho_{a,34} = -i \sum_{i,j,k=0}^{\infty} c_{i+1,j,k+1,i+1,j,k}(0,\theta,x) \sin^2(g\tau\sqrt{i+1}) \cos^2(g\tau\sqrt{j}) \sin(g\tau\sqrt{k+1}) \cos(g\tau\sqrt{k})$$

$$\rho_{a,35} = \sum_{i,j,k=0}^{\infty} c_{i+1,j,k+1,i,j+1,k+1}(0,\theta,x) \sin(g\tau\sqrt{i+1}) \cos(g\tau\sqrt{i}) \sin(g\tau\sqrt{j+1}) \cos(g\tau\sqrt{j}) \sin^2(g\tau\sqrt{k+1})$$

$$\rho_{a,36} = -i \sum_{i,j,k=0}^{\infty} c_{i+1,j,k+1,i,j+1,k}(0,\theta,x) \sin(g\tau\sqrt{i+1}) \cos(g\tau\sqrt{i}) \sin(g\tau\sqrt{j+1}) \cos(g\tau\sqrt{j}) \sin(g\tau\sqrt{k+1}) \cos(g\tau\sqrt{k})$$

$$\rho_{a,37} = -i \sum_{i,j,k=0}^{\infty} c_{i+1,j,k+1,i,j,k+1}(0,\theta,x) \sin(g\tau\sqrt{i+1}) \cos(g\tau\sqrt{i}) \cos^2(g\tau\sqrt{j}) \sin^2(g\tau\sqrt{k+1})$$

$$\rho_{a,38} = - \sum_{i,j,k=0}^{\infty} c_{i+1,j,k+1,i,j,k}(0,\theta,x) \sin(g\tau\sqrt{i+1}) \cos(g\tau\sqrt{i}) \cos^2(g\tau\sqrt{j}) \sin(g\tau\sqrt{k+1}) \cos(g\tau\sqrt{k})$$

(A2)

$$\rho_{a,45} = i \sum_{i,j,k=0}^{\infty} c_{i+1,j,k,i,j+1,k+1}(0,\theta,x) \sin(g\tau\sqrt{i+1}) \cos(g\tau\sqrt{i}) \cos(g\tau\sqrt{j}) \sin(g\tau\sqrt{j+1}) \sin(g\tau\sqrt{k+1}) \cos(g\tau\sqrt{k})$$

$$\rho_{a,46} = \sum_{i,j,k=0}^{\infty} c_{i+1,j,k,i,j+1,k}(0,\theta,x) \sin(g\tau\sqrt{i+1}) \cos(g\tau\sqrt{i}) \cos(g\tau\sqrt{j}) \sin(g\tau\sqrt{j+1}) \cos^2(g\tau\sqrt{k})$$

$$\rho_{a,47} = \sum_{i,j,k=0}^{\infty} c_{i+1,j,k,i,j,k+1}(0,\theta,x) \sin(g\tau\sqrt{i+1}) \cos(g\tau\sqrt{i}) \cos^2(g\tau\sqrt{j}) \sin(g\tau\sqrt{k+1}) \cos(g\tau\sqrt{k})$$

$$\rho_{a,48} = -i \sum_{i,j,k=0}^{\infty} c_{i+1,j,k,i,j,k}(0,\theta,x) \sin(g\tau\sqrt{i+1}) \cos(g\tau\sqrt{i}) \cos^2(g\tau\sqrt{j}) \cos^2(g\tau\sqrt{k})$$

$$\rho_{a,56} = -i \sum_{i,j,k=0}^{\infty} c_{i,j+1,k+1,i,j+1,k}(0,\theta,x) \cos^2(g\tau\sqrt{i}) \sin^2(g\tau\sqrt{j+1}) \sin(g\tau\sqrt{k+1}) \cos(g\tau\sqrt{k})$$

$$\rho_{a,57} = -i \sum_{i,j,k=0}^{\infty} c_{i,j+1,k+1,i,j,k+1}(0,\theta,x) \cos^2(g\tau\sqrt{i}) \sin(g\tau\sqrt{j+1}) \cos(g\tau\sqrt{j}) \sin^2(g\tau\sqrt{k+1})$$

$$\rho_{a,58} = - \sum_{i,j,k=0}^{\infty} c_{i,j+1,k+1,i,j,k}(0,\theta,x) \cos^2(g\tau\sqrt{i}) \sin(g\tau\sqrt{j+1}) \cos(g\tau\sqrt{j}) \sin(g\tau\sqrt{k+1}) \cos(g\tau\sqrt{k})$$

$$\rho_{a,67} = \sum_{i,j,k=0}^{\infty} c_{i,j+1,k,i,j,k+1}(0,\theta,x) \cos^2(g\tau\sqrt{i}) \sin(g\tau\sqrt{j+1}) \cos(g\tau\sqrt{j}) \sin(g\tau\sqrt{k+1}) \cos(g\tau\sqrt{k})$$

$$\rho_{a,68} = -i \sum_{i,j,k=0}^{\infty} c_{i,j+1,k,i,j,k}(0,\theta,x) \cos^2(g\tau\sqrt{i}) \sin(g\tau\sqrt{j+1}) \cos(g\tau\sqrt{j}) \cos^2(g\tau\sqrt{k})$$

$$\rho_{a,78} = -i \sum_{i,j,k=0}^{\infty} c_{i,j,k+1,i,j,k}(0,\theta,x) \cos^2(g\tau\sqrt{i}) \cos^2(g\tau\sqrt{j}) \sin(g\tau\sqrt{k+1}) \cos(g\tau\sqrt{k}) \tag{A3}$$

### APPENDIX B: CV FIELD APPROXIMATED BY A THREE QUBIT MIXED STATE

Here we report some results related to the case of CV field prepared in the mixed state of Eq. (19). From the atomic density matrix elements $\rho_a$ reported in Eq. (20) we can derive the following expression of the atomic purity $\mu_a^{(M)}(g\tau,p)$:

$$\begin{aligned}
\mu_a^{(M)}(g\tau,p) &= \frac{1}{4}\Bigg\{ \Big[ p + 2(1-p)\cos^2(g\tau) + p\cos^6(g\tau) \Big]^2 + \sin^4(g\tau)\Big[4(1-p) + 3p\cos^4(g\tau)\Big] + \\
&\quad + p^2 \sin^6(g\tau)\Big[ 2 + 3\cos^4(g\tau)\sin(g\tau)^2 + \sin^6(g\tau) \Big] \Bigg\}
\end{aligned} \tag{B1}$$

For the partial transpose matrices $\rho_a^{tI}$ ($I = A, B, C$) we find that they all have the same two eigenvalues $\lambda_1^-$ and $\lambda_2^-$ that can assume negative values for some $p$ and $g\tau$:

$$\begin{aligned}
\lambda_1^-(g\tau,p) &= \frac{\sin^2(g\tau)}{12}\Big\{ 2(1-p) + 3p\cos^2(g\tau) - \Big[ (2(1-p) + 3p\cos^2(g\tau))^2 + 12p\sin^2(g\tau) \times \\
&\quad \times [3p(1 - p\cos^6(g\tau)) - 2(1-p)\cos^2(g\tau)] \Big]^{\frac{1}{2}} \Big\} \\
\lambda_2^-(g\tau,p) &= \frac{1}{12}\Big\{ 3\Big[ p + 2(1-p)\cos^2(g\tau) + p\cos^2(\cos^4(g\tau) + \sin^4(g\tau)) \Big] + \\
&\quad - \Big[ 9\Big( p + 2(1-p)\cos^2(g\tau) + p\cos^2(\cos^4(g\tau) + \sin^4(g\tau)) \Big)^2 +, \\
&\quad + 4\sin^4(g\tau)\Big( 8(1-p)^2 - 9p^2\cos^2(g\tau)(1 + \cos^6(g\tau)) - 18p(1-p)\cos^4(g\tau) \Big) \Big]^{\frac{1}{2}} \Big\}
\end{aligned} \tag{B2}$$

From the above eigenvalues we can evaluate numerically the tripartite negativity as $N_{ABC}^{(M)} = -2(\lambda_1^- + \lambda_2^-)$ for negative values of $\lambda_1^-$ and $\lambda_2^-$ and zero elsewhere.